\def\be{\begin{eqnarray}}
\def\ee{\end{eqnarray}}
\def\bea{\begin{eqnarray}}
\def\eea{\end{eqnarray}}
\def\beas{\begin{eqnarray*}}
\def\eeas{\end{eqnarray*}}
\newcommand{\eq}[1]{Eq.~(\ref{#1})}

\def\bfb{{\bf b}}

\def\bfR{{\bf R}}

\def\bfq {{\bf q}}

\def\bfb{{\bf b}}
\def\bfB{{\bf B}}
\def\bfk{{\bf k}}

\def\bfp{{\bf p}}

\documentclass{ws-ijmpe}

\begin{document}

\markboth{MILLER and ARRINGTON}{Inclusive Exclusive Connection}

\catchline{}{}{}{}{}

\title{THE INCLUSIVE-EXCLUSIVE CONNECTION AND THE NEUTRON
NEGATIVE CENTRAL CHARGE DENSITY}

\author{              GERALD A. MILLER}

\address{Physics    Department, University of Washington\\
Seattle, WA, 98195-1560,
USA\\
miller@phys.washington.edu}

\author{JOHN ARRINGTON}

\address{ Physics Division, Argonne National Laboratory, \\
Argonne, IL, 60439\\
johna@anl.gov}

\maketitle

\begin{history}
\received{(received date)}
\revised{(revised date)}
\end{history}

\begin{abstract}
We find an interpretation of the recent finding that the central charge density of the neutron is
negative by using
models of generalized parton distributions at zero skewness to relate
the behavior of deep inelastic scattering quark distributions, evaluated at
high $x$, to the transverse charge density evaluated at small
distances. The key physical input of these models is the
Drell-Yan-West relation  We find that 
the $d$ quarks dominate the neutron structure
function for large values of Bjorken $x$, where the large longitudinal momentum of the
struck quark has a significant impact on determining the center-of-momentum of the system,
and thus the ``center'' of the nucleon in the transverse position plane.
\end{abstract}

\section{ Outline}

Electron scattering is the preferred tool for extracting information on the
spatial and momentum distribution of the quarks in nucleons.  High energy
scattering provides a clean picture of the quarks' momentum distribution
for a nucleon boosted into infinite momentum frame (IMF).  Measurements of
elastic scattering at lower energy scales allow extraction of the nucleon form
factors, which can be related to the spatial distribution of charge in the 
rest frame of the nucleon.  However, obtaining rest frame charge distributions
requires model-dependent relativistic boost corrections, limiting our ability
to extract these distributions from data.  Significant work has gone into 
better understanding the issues involved in studying nucleon distributions, as
well as providing unified descriptions of the space and momentum distributions
of the quarks.

The present discussion and the papers on which it is based would not have been
possible without the great amount of experimental technique, effort and
ingenuity that has been used recently to measure the electromagnetic form
factors of the nucleon\cite{gao03,hydewright04, perdrisat07,arrington07a}.
These quantities are probability amplitudes that the nucleon can absorb a
given amount of momentum and remain in the ground state, and are related to
the nucleon charge and magnetization densities.

We note that there will be an Institute for Nuclear Theory Program 
held in the Fall of 2009 that is devoted to the electromagnetic physics of the
Jefferson Laboratory upgrade to 12 GeV.  Please see the website:
www.int.washington.edu/PROGRAMS/09-03.html.

 We begin our analysis by reviewing recent
work\cite{miller07} that determines the transverse charge density of the
neutron in a model independent way. This work surprisingly found that the
central charge density is negative. Then we discuss how the
inclusive-exclusive connection is used to provide an interpretation for this
fact\cite{ma08}.

\section{Definitions and the density interpretation}

 We begin by  by presenting    definitions of the  form
factors. Let $J^\mu(x^\nu)$ be the electromagnetic current operator, in units of the proton 
charge. Then  the nucleon form factors are given by the matrix element
\bea
\langle p',\lambda'| J^\mu(0)| p,\lambda\rangle =\bar{u}(p',\lambda')
\left(\gamma^\mu F_1(Q^2)+i\frac{\sigma^{\mu\alpha}}{2M}q_\alpha F_2(Q^2)
\right) u(p,\lambda),\eea
where $M$ is the nucleon mass, and  the momentum transfer $q_\alpha=p'_\alpha-p_\alpha$ is taken as space-like, so that 
$Q^2\equiv -q^2>0.$ The nucleon polarization states are chosen to be those of 
definite light-cone helicities $\lambda,\lambda'$.\cite{soper}  
The charge (Dirac) form factor is $F_1$, normalized such that $F_1(0)$ is the
nucleon charge, and the magnetic (Pauli) form factor is $F_2$, normalized such that $F_2(0)$ is the
anomalous magnetic moment. The Sachs form factors  are
$
G_E(Q^2)\equiv F_1(Q^2)-\frac{Q^2}{4M^2}F_2(Q^2),\; G_M(Q^2)\equiv F_1(Q^2)+F_2(Q^2).$

In  the Breit frame, in which 
$\bfp=-\bfp'$,  $G_E$ is the nucleon helicity flip matrix element of
$J^0$. This has been interpreted as meaning that $G_E$ is the
three-dimensional
Fourier transform of the charge density in the rest frame. Indeed, 
 the scattering of neutrons from the electron cloud of atoms  measures
the derivative $-\frac{1}{6}dG_E(Q^2)/dQ^2$ at $Q^2=0$, which has  
been widely interpreted as
the mean-square charge radius of the neutron.  However,
a direct probability or density interpretation of $G_E$ is
spoiled by a non-zero  value of $Q^2$, no matter how small. This is because 
relativistic dynamics, which cause
the wave functions of the initial and final nucleons of different
momenta to differ, must be used. The final wave function is related
to the initial one by a complicated  boost operator that when acting
on the initial state of a given momentum changes it to the same of the
different final momentum.  In general the boost operator (or boost)
contains the effects of interactions. Thus the  initial and final
states differ, invalidating  a probability or density interpretation.
It is only in the case that non-relativistic dynamics are applicable that form
factors are simply the Fourier transforms of the rest frame spatial
distributions.

It has commonly been assumed that at low momentum transfers, these corrections
could be safely neglected.  If we treat relativistic corrections as in atomic
physics, they are governed by $ v^2/c^2\sim p^2/m^2$, where $m$ is the
mass of the boosted constituent.  Computing the form factors involves
integrating over the momentum $p$ so that it is replaced by the momentum
transfer $Q$.  Thus for small values of $Q^2$ we find \bea G_E(Q^2)
=G_E(0)-
{ Q^2\over6}
\left(\int d^3r r^2 \rho(r) + C/m^2\right),\eea where $C$ is an unknown
coefficient. 
 In a simple constituent quark model, $m \approx 300$~MeV, and so
these $1/m^2$ boost corrections are negligible only for $p^2 \ll
 0.1$~GeV$^2$.
However, for the neutron $G_E(0)=0$, so the corrections are expected to
 be relatively  large. 
The boost correction term need not be small compared to the finite size
contribution unless $C\ll 1$; since both corrections scale with $Q^2$, 
the relative correction to the extracted radius does not
vanish as $Q^2$ approaches 0.
 The constituent
quarks represent the low $Q^2$, dressed versions of the near-massless current
quarks of QCD. 
For current quarks of mass 5--10~MeV, the boost corrections are
important for all $Q^2$ values where measurements exist.  While it is possible
to construct systematically improveable models which should yield a complete
description of the nucleon, it is not clear how one would quantitatively
determine how well any particular model fully reproduces the corrections
associated with the relativistic boost.  Thus, there is always some model
dependence in the extraction of the rest frame charge distributions from
the form factors, and it is not  clear how well one can quantify these
corrections and uncertainties, even at very low $Q^2$.

\section{Light Cone Coordinates, the  Infinite Momentum Frame, and the
Drell-Yan Frame}

The use of light cone coordinates and the kinematic subgroup of the
Poincar\'{e} group, which is closely related to the use
of the infinite momentum frame, enables one to avoid the difficulties
associated with including the boost.

The basic idea is that the ``time'' variable is given by
$ x^+=(ct+z)/\sqrt{2} =(x^0+x^3)/\sqrt{2}$ 
and the conjugate
evolution operator is given by 
$ p^-=(p^0-p^3)/\sqrt{2}.$ 
Starting in an ordinary reference
frame and making a Lorentz transformation into a frame moving with
nearly the speed of light in the $3$ direction converts the usual $t$
into $x^+$. The 3 spatial variables must be different than $x^+$, so
we take $
x^-=(x^0-x^3)/\sqrt{2}.$ 
If $x^+=0$ then $x^-=-\sqrt{2}z$ and
$x^-$ can be thought of as something like a $z$ or $x^3$ variable, but
rotational invariance can not generally be used to relate  the $x^-$
and $x,y$ dependence of the density. The canonically conjugate
momentum to $x^-$ is $
p^+=(p^0+p^3)/\sqrt{2}. $ 
For the transverse degrees of freedom
we use the usual position $(x,y)$ and momentum $(p_x,p_y)$ variables denoted by
$\bfb,\bfp$ where the boldface notation denotes a transverse vector. 

We also exploit the kinematic subgroup. That is transverse Lorentz
transformations, with a transverse velocity  ${\bf v}$
(or boosts in the transverse $x,y$ or $\bfb$ direction) do not involve
interactions.
In particular,  these transformations are defined by
$k^+\rightarrow k^+,\bfk\rightarrow \bfk-k^+{\bf v},$ 
with 
$k^-$ changed so that $k_\mu k^\mu$ is not changed. These
transformations are just like the non-relativistic Galilei
transformation except that $k^+$ appears instead of a mass. This means
that we are allowed to do Fourier transformations of variables provided
transverse degrees of freedom are involved.

If the momentum transfer vector $q^\mu$ is space-like we may use the
so-called Drell-Yan (DY) frame in which $q^+=0$. This means that 
the plus component of the momentum of the nucleon is the same before
and after the absorption of the single photon. Then the momentum
transfer vector is in the transverse direction, so that usual
two-dimensional Fourier
transform techniques can be used.

\section{Charge density of the neutron}
A proper determination of a 
 charge density requires the measurement of a density operator.
We shall show that measurements of the pion form factor directly involve the 
three-dimensional parton charge
density operator, in the infinite momentum frame IMF,  
$\hat{\rho}_\infty(x^-,\bfb)=J^+(x^-,\bfb)$.
In this frame  the electromagnetic charge density 
$J^0$ becomes $J^+$ and
\bea \hat{\rho}_\infty(x^-,\bfb)
=\sum_q e_q \overline{q}(x^-,\bfb)\gamma^+q(x^-,\bfb)=\sum_q e_q \sqrt{2} 
q^\dagger_+(x^-,\bfb)q_+(x^-,\bfb),\label{imfop}\eea
where 
 $q_+(x^\mu)=
\gamma^0\gamma^+/\sqrt{2} q(x^\mu)$, the independent part of the
 quark-field operator $q(x^\mu)$.
We set the 
  time variable, $x^+=(t+z)/\sqrt{2},$ to  zero, and do not display
 it in  any function.

 The spatial structure of a  hadron can be examined if
one uses\cite{soper77,mbimpact,diehl2} 
states that are transversely localized.  The state with transverse center of mass
$\bfR$ set to {\bf 0} is formed by taking a  linear superposition of
states of transverse momentum:
\be 
\left|p^+,{\bf R}= {\bf 0},
\lambda\right\rangle
\equiv {\cal N}\int \frac{d^2{\bf p}}{(2\pi)^2} 
\left|p^+,{\bf p}, \lambda \right\rangle,
\label{eq:loc}
\ee
where $\left|p^+,{\bf p}, \lambda \right\rangle$
are light-cone helicity eigenstates
\cite{soper} and
${\cal N}$ is a normalization factor satisfying
$\left|{\cal N}\right|^2\int \frac{d^2{\bf p}_\perp}{(2\pi)^2}=1$. 
The expansion \eq{eq:loc} makes sense only in the infinite momentum
frame $p^+\rightarrow\infty$ because we must have $2p^+p^--\bfp^2=M^2$.
The
nucleon states are normalized as
$\langle {p'}^+,\bfp'\lambda'| {p}^+,\bfp,\lambda\rangle
=2p^+(2\pi)^3 \delta_{\lambda',\lambda} \delta({p'}^+-p^+)\delta^{(2)}({\bfp}'-\bfp)$.

Next  we relate the charge density  
\bea {\rho}_\infty(x^-,\bfb)=
\frac{ \left\langle p^+,{\bf R}= {\bf 0},
\lambda\right| \hat{\rho}_\infty(x^-,\bfb)
\left|p^+,{\bf R}= {\bf 0},
\lambda\right\rangle}
{\left\langle p^+,{\bf R}= {\bf 0},
\lambda|p^+,{\bf R}= {\bf 0},
\lambda\right\rangle},\eea
to $F_1(Q^2)$.
In the DY frame no momentum is transferred in the plus-direction, so  that
information regarding the $x^-$ dependence of the distribution is not
accessible. 
Therefore we 
integrate over $x^-$, using the relationship translational invariance,
and then use our momentum expansion and the definition of $F_1$ as the
helicity non-flip matrix element of $J^+$ in a Drell-Yan frame
 to find
\bea&&
\rho(b)\equiv\int \;dx^-\rho_\infty(x^-,b)\label{rhobtx}\\
&&\rho(b)=\int \frac{d^2q}{(2\pi)^2} F_1(Q^2=\bfq^2)
e^{-i\bfq\cdot\bfb}=\int \frac{Q\; dQ}{(2\pi)} F_1(Q^2)J_0(Qb)
e^{-i\bfq\cdot\bfb}, 
\label{rhob0}\eea
where $J_0$ is a cylindrical Bessel function and $\rho(b)$ is termed the
transverse charge density, giving the charge density at a transverse position
$b$, integrating over all longitudinal momentum. In the infinite momentum
frame, the longitudinal dimension of the nucleon is contracted to a point,
and only the transverse position remains.  The value of $b=0$ corresponds to
the center of longitudinal-momentum of the nucleon in the transverse
dimension.

It is not as straightforward to isolate the magnetization distribution,
and there have been multiple such densities proposed \cite{miller07b,carlson08}.
Starting with the matrix element of ${\bf J}\cdot {\bf A}$ and working in the infinite
momentum frame leads to the result that 
the transverse magnetization density to
is  the  two-dimensional Fourier transform of $F_2$, just as the charge
density is the transform of $F_1$ \cite{miller07b}.
This interpretation yields a difference between the magnetic and electric
radii in the proton.

\begin{figure}
\unitlength1.cm
\begin{picture}(14,10.8)(1.8,1.1)
\includegraphics{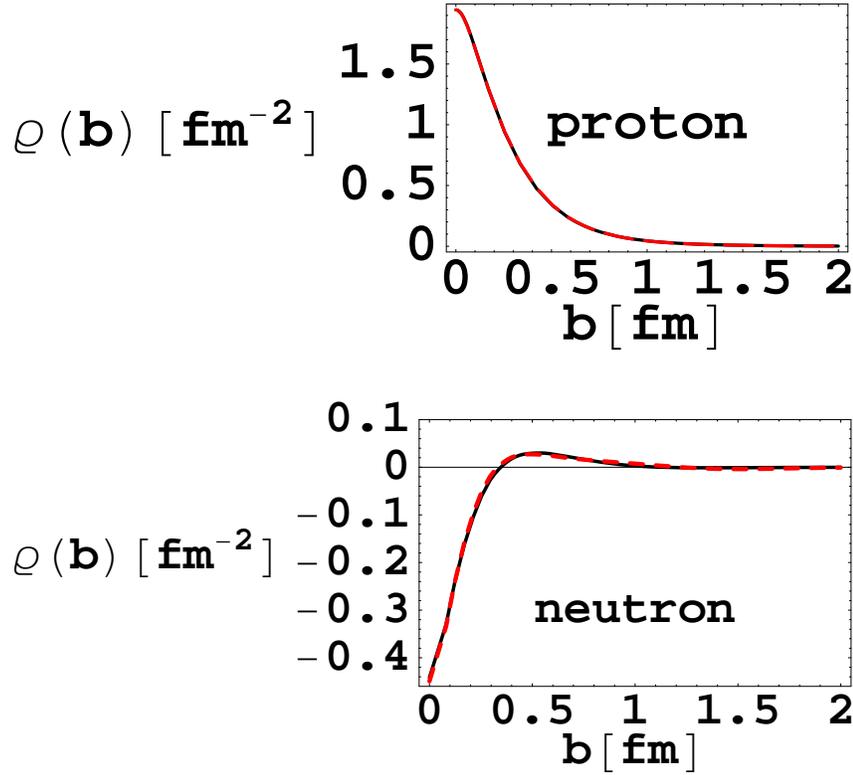}
\end{picture}\label{proton}
\caption{Upper panel: proton transverse
charge density $\rho(b)$. Lower panel: neutron
transverse charge density. The solid curves use the parameterization of Kelly, and the
dashed (red) curves use Bradford, \textit{et al.}. Reprinted with
permission from Ref.~$^5$.
Copyright 2007 by the American Physical Society.}
\label{fig:proton}
\end{figure}

Here we
exploit \eq{rhob0} by  using  recent parameterizations 
\cite{Bradford:2006yz,Kelly:2004hm,Arrington:2003qk} of measured form factors
to determine $\rho(b)$. 
Applying  \eq{rhob0} to the proton 
using two sets of form factor parameterizations\cite{Bradford:2006yz,Kelly:2004hm}
yields the results shown in the upper panel of Fig.~\ref{fig:proton}. The curves
obtained using the two different parameterizations overlap. 
Furthermore, there is
negligible sensitivity to form factors at very high values of $Q^2$ that are currently unmeasured. The density is peaked 
at low  values of $b$, but has a long positive tail, suggestive of a long-ranged, positively charged 
pion cloud.

The neutron  results for $\rho(b)$ are shown in the lower panel of
Fig.~\ref{fig:proton}. The curves obtained using the two different
parameterizations seem to overlap, but we will return to this below. The
surprising result is that the central neutron charge density is negative.
The negative nature of the neutron's central charge density appears
to contradict two current ideas. If the neutron is sometimes a proton
surrounded by a negatively charged pionic cloud, one would expect to obtain a
positive central density\cite{cbm}. Another  mechanism involving correlations
in the nucleonic wave function induced by  one gluon exchange would also  lead
to a positive central density because  the interaction between two identical
$d$ quarks\cite{isgurk} is repulsive.

\begin{center}
 \begin{figure}[htb]
   \includegraphics[width=10.0cm,height=5.5cm]{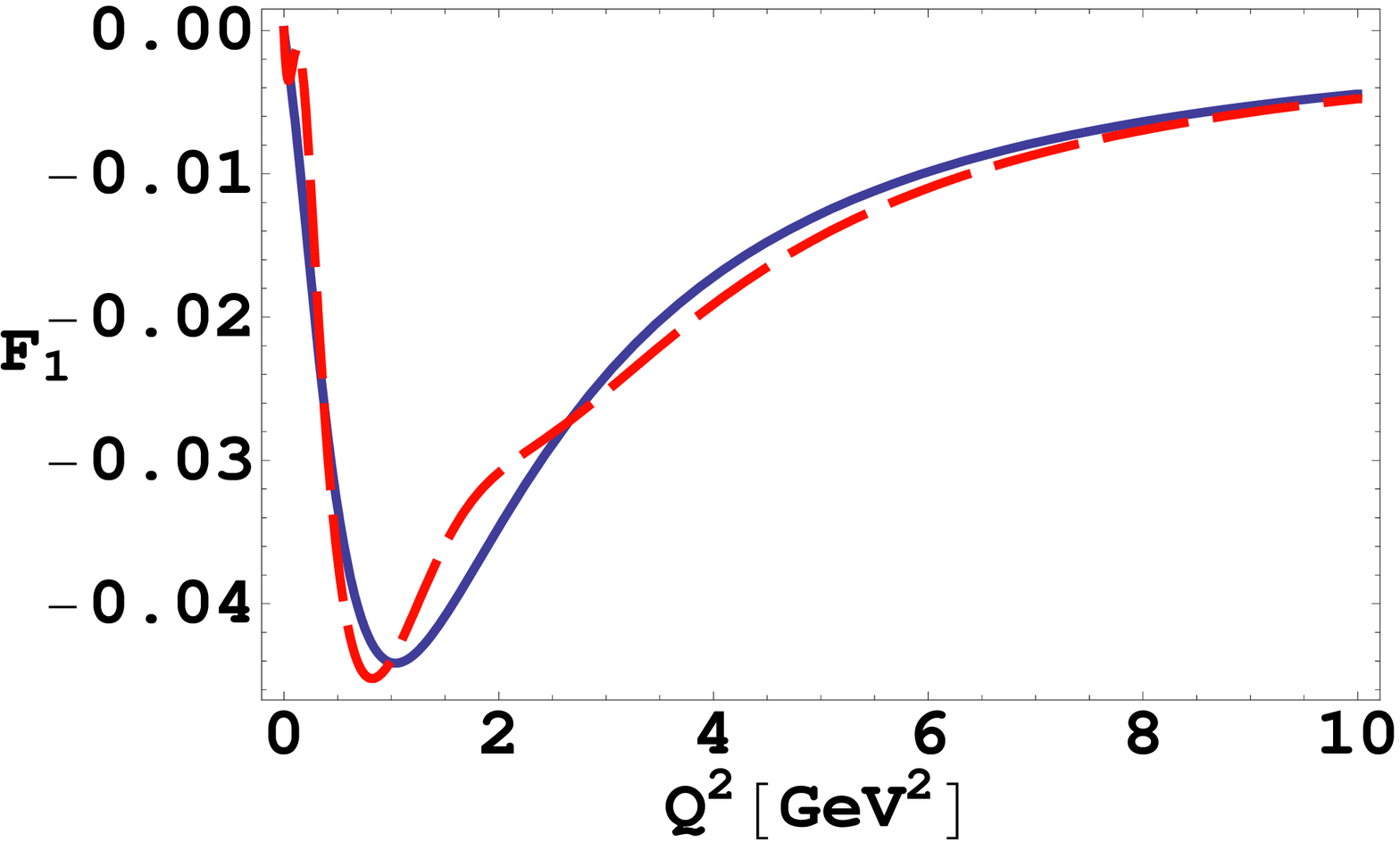}
   \includegraphics[width=10.0cm,height=5.5cm]{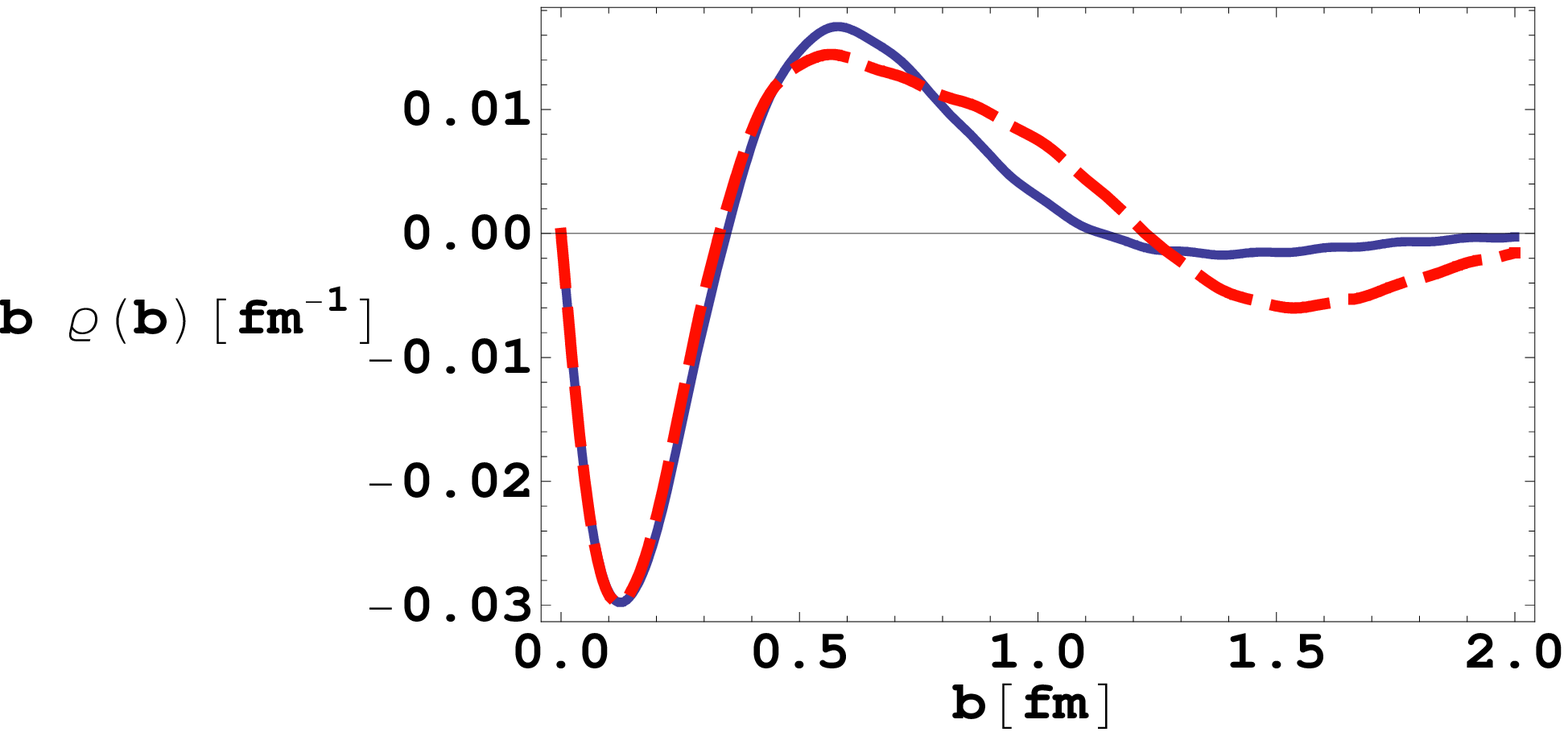}
   \caption{Neutron $F_1$ and transverse charge density. Upper panel: $F_1$. Lower panel:   $b \rho(b)$ in transverse position space. The
solid curves are obtained using the fits of Kelly, and the dashed curves
the fits of Bradford, \textit{et al.}. Reprinted with permission from Ref.~$^5$. Copyright 2007 by the American Physical Society.}
   \label{fig:f1}
 \end{figure}
\end{center}

The resultant negative central density thus deserves further examination. 
The upper panel of Fig.~\ref{fig:f1} shows $F_1$ for the neutron obtained
using the two different parameterizations\cite{Kelly:2004hm,Bradford:2006yz}
which are observably different. However, in both cases.  $F_1$ is  negative
for all  values of $Q^2$. If $F_1$ is always negative, then taking
$b=0,\;J_0(Qb)=1$ in \eq{rhob0}, will always yield a negative central density.
The long range structure of the charge density is captured by displaying the
quantity $b\rho(b)$ in the lower panel of Fig.~\ref{fig:f1}. At very large
distances from the center, the charge density is negative, as expected in the
pion cloud picture.

These findings appear to contradict previous understanding of the nucleon
charge distributions based on the model-dependent extraction of the rest frame
charge distributions.  Because of the large and model-dependent boost
corrections at large $Q^2$, one cannot obtain information about the rest frame
charge density at the very center of the nucleon from measurements of the
form factors.  While there is no direct experimental information on this
possibility of a very small negative core in the rest frame distribution, this
negative core seems to contradict
 the accepted explanations of the origin of the long
range negative cloud  which occurs along with a positively charged interior.
In addition, the negative core in the IMF transverse
density is a feature even in models that  build up the charge
distribution based on a pion cloud model.  It is clearly important to
understand the differences between the IMF charge density and the rest frame
charge density to fully understand the new features of these model-independent
spatial distributions.

The surprising model independent result is that the
density of the neutron is negative. The remainder of this presentation
is concerned with trying to explain this remarkable feature of nature.

\section{Inclusive-exclusive connection}


Generalized parton distributions (GPDs) contain
information about the longitudinal momentum fraction $x$ as well as the
transverse position $b$.  Information regarding the $x$ and $b$ dependence is
obtained from experiment by using GPDs to reproduce both deep inelastic
scattering and elastic scattering data. Thus we use this inclusive-exclusive
connection to better understand the central neutron charge density.

The widely studied GPDs\cite{ji96,radyushkin97} are of high current interest
because they can be related to the total angular momentum carried by quarks in
the nucleon. We consider the specific case in which the longitudinal momentum
transfer $\xi$ is zero, and the initial and final nucleon helicities are
identical ($\lambda'=\lambda$). Then, in the light-cone gauge, $A^+=0$, the
matrix element defining the GPD $H_q$ for a quark of flavor $q$ and zero
skewness is
\be
H_q(x,t) = \langle p^+,\bfp',\lambda|\widehat{O}_q(x,{\bf 0})
|p^+,\bfp,\lambda\rangle,
\label{eq:pd}
\eea
where
\be
\widehat{O}_q(x,{\bf b}) \equiv
\int \frac{dx^-}{4\pi}{q}_+^\dagger
\left(-\frac{x^-}{2},{\bf b} \right) 
q_+\left(\frac{x^-}{2},{\bf b}\right) 
e^{ixp^+x^-}.
\label{eq:bperp}
\ee
We abbreviate $H_q(x,\xi$=$0,t)\equiv H_q(x,t)$ and $
-t=-(p'-p)^2=(\bfp'-\bfp)^2=Q^2.$ 

It is well known that GPDs provide a unified description of a number of
hadronic properties.\cite{ji96}  Of particular interest here is that for
$t$=0 they reduce to conventional PDFs, $H_q(x,0)= q(x)$, and that the
integration of the charge-weighted $H_q$ over $x$ yields the nucleon
electromagnetic form factor:
\be
F_1(t)=\sum_q e_q \int dx H_q(x,t).  
\label{eq:form}
\ee

The impact parameter-dependent PDF\cite{burkardt00} for a quark of
flavor $q$ is the matrix element
of the operator $\widehat {O}_q$ in the state $\left|p^+,{\bf R}= {\bf 0},
\lambda\right\rangle$:
\be
\rho_\perp^q({\bf b},x) \equiv 
\left\langle p^+,{\bf R}= {\bf 0},
\lambda\right|
\widehat{O}_q(x,{\bf b})
\left|p^+,{\bf R}= {\bf 0},
\lambda\right\rangle. 
\label{eq:def1}
\ee
We use the notation $\rho_\perp^q({\bf b},x)$ instead of the originally
defined\cite{burkardt00} $q(x,\bfb)$ because this quantity  is a
density that  gives the probability that the quark has  a longitudinal momentum
fraction $x$ and is at a transverse position $\bfb$.
The quantity $\rho_\perp^q({\bf b},x)$ is the two-dimensional Fourier
transform of the GPD $H_q$:
\bea
\rho_\perp^q({\bf  b},x)=\int \frac{d^2q}
{(2\pi)^2}e^{-i\;\bfq\cdot\bfb}H_q(x,t=-\bfq^2).\label{ft1}
\eea
%

We extract the form factor $F_1$ by integrating
$\rho_\perp^q({\bf b},x)$ over all values of $x$, multiplying by the quark
charge $e_q$, and summing over quark flavors $q$.\cite{soper77} The result is
the IMF charge density in transverse space:
\bea
&\rho(b) & \equiv \sum_q e_q\int dx\;\rho_\perp^q({\bf b},x)  .
\label{rhobt}
\eea
The relations \eq{rhobt} and \eq{rhobtx} provide two expressions for
the
transverse density $\rho(b)$. The impact parameter GPD and the
three-dimensional
density are related by Parseval's theorem.
The quantity $\rho(b)$ gives the charge density at a transverse position $\bfb$
irrespective of the longitudinal momentum fraction or longitudinal position. 

There is a tight connection between the values of
$x$ and the values of $\bfb$. For a given Fock space component, the
center of transverse momentum (CM) $\bfR$ is given by
\bea \bfR=0=\sum_ix_i\bfb_i, \label{tcm}\eea
where $x_i,\bfb_i$ are the longitudinal momentum fractions and
transverse positions of the $i$'th partons.  Using this definition, the
longitudinal momentum of a quark determines it's impact on defining the
transverse position of the nucleon.  If the struck quark, $x_i$,
carries nearly all of the plus component of the total momentum, then
the other quarks must carry a total plus momentum of $1-x_i$ which approaches
zero as $x_i \to 1$.  Thus, $x_j \to 0$ for $i \ne j$, and via \eq{tcm},
$\bfb_i$ must also approach zero as long as the values of $\bfb_j$ remain 
finite. Thus, large values of $x$ correspond to a small value of $b$, as
the struck quark plays a large role in defining the transverse center of
mass as $x \to 1$.

Let us investigate $\rho_\perp(\bfb,x)$ to understand the origin of the
neutron's negative central charge density. The quantities are not measured
directly, but have been obtained from models that incorporate fits to parton
distributions and electromagnetic nucleon form
factors.\cite{guidal05,diehl05,ahmad07,tiburzi04} This method exploits form
factor sum rules at zero skewness to model
information regarding the valence quark GPDs, $H^q_v\equiv H^q-H^{\bar{q}}$.
This yields the net contribution to the form factors from quarks and
anti-quarks, although it does  not correspond to the valence distribution
within a model for which   sea distributions for quarks and antiquarks have
different $x$ or $t$ dependences. The effects of strangeness are
neglected in these fits.

Each parameterization we use\cite{guidal05,diehl05,ahmad07} incorporates 
the Drell-Yan-West\cite{DYW} relationship between the behavior of the
structure function $\nu W_2(x)$ function near $x=1$, measured in
inclusive reactions  and the behavior of the electromagnetic
form factor at large values of $Q^2$, measured in the exclusive
elastic scattering process. In particular, for a system of
$n+1$ valence quarks, described by a power-law wave function
\bea
\lim_{x\rightarrow1}\nu W_2(x)=(1-x)^{2n-1} \to
\lim_{Q^2\rightarrow\infty}F_1(Q^2)=\frac{1}{Q^{2n}},
\eea   
with the relation being that the same value of $n$ that defines the high-$x$
behavior of the structure function also defines the high-$Q^2$ behavior
of the form factor, thus associating the behavior of large values of $x$
with large momentum transfers, $Q^2=\bfq^2$, which in turn correspond to small
values of $b$.  So again we see we see a connection between large values of
$x$ and small values of $b$.

To proceed further we use specific forms of the GPDs,
and these determine the details of the results. Diehl {\it et
al.}\cite{diehl05} use
\bea
H^q_v(x,t)=q_v(x) \exp[f_q(x)t],
\eea
where 
\bea
f_q(x)=[\alpha'\log[1/x]+B_q](1-x)^3+A_qx(1-x)^2,
\eea
is the form that gives the best fit to the data. The parameter $\alpha'$
represents the slope of the Regge trajectory, and the
CTEQ6 PDFs\cite{pumplin02} are taken as input. Here we use the best fit
parameters, taken from the second line of Table 8
of Ref.~\cite{diehl05}. These are
detailed in Ref.~\cite{ma08}.
The labels $q$ refer to $u$ and $d$,
the $u$ and $d$ quarks in the proton. These correspond to $d$ and $u$ quarks
in the neutron, if charge symmetry~
\cite{miller90,miller98,londergan98,miller06} 
is upheld.  For the proton, $2d_v/u_v$ falls
rapidly for large values of $x$, which means that $u$ quarks dominate the
parton distribution for large  values of $x$. For the neutron, the assumption
of charge symmetry implies that $d_v$ for the neutron is the same as $u_v$ for
the proton, and vice-versa. Thus, the $d$ quarks dominate the parton
distribution in the neutron for large values of $x$. The
distributions of Ref.~\cite{guidal05} have $A_q=B_q=0$ and
$f_q(x)=[\alpha'_q\log[1/x]](1-x).$ Those of Ref.~\cite{ahmad07} have a more
complicated form and include the additional  constraint that the nucleon consists of
three quarks at an initial scale of $Q_0^2=0.094$ GeV$^2$.

We study  the connection between regions of $x$ and regions
of $b$. To do this define
\bea
\rho_\perp^q(b,\Delta x)\equiv \int_{\Delta x} dx~e_q~\rho_\perp^q(b,x),
\eea
where $e_q$ is the quark charge in units of the proton charge 
($e_u=2/3,\;e_d=-1/3$)
with $\rho_\perp^{p,n}$ being obtained from
appropriate sums of $\rho_\perp^{q}$.  This represents the contribution to the
charge density from quarks in the $x$ region defined by $\Delta x$, rather
than the total density obtained by integrating over all $x$.

\begin{center}
 \begin{figure}[htb]
   \includegraphics[width=8.5cm]{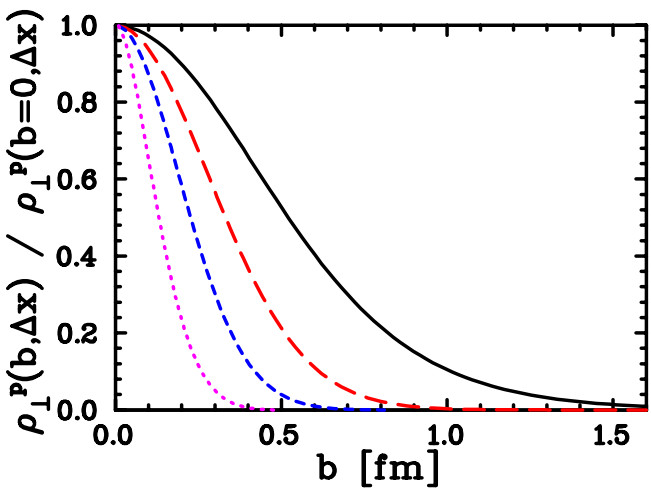}
   \caption{  
The proton transverse charge density, $\rho_\perp^p(b,\Delta x)$, for quarks in
different $\Delta x$ regions: $x$$<$0.15 (solid), 0.15$<$$x$$<$0.3 (long-dash),
0.3$<$$x$$<$0.5 (short-dash), and $x$$>$0.5 (dotted).  The curves have been
normalized to unity at $b=0$ to emphasize the variation in
   width. Reprinted with permission from Ref.~$^6$.
 Copyright 2008 by the American Physical Society.}
     \label{fig:proton_scaled}
 \end{figure}
\end{center}

We have pointed out above that at large $x$, the struck quark plays a
significant role in defining the transverse CM, so the distribution of
high-$x$ quarks becomes localized at small values of $b$.  This is clearly
visible in Fig.~\ref{fig:proton_scaled}, which shows $\rho_\perp^p(b,\Delta x)$
for different bins in $\Delta x$. The curves have been scaled to yield unity
at $b=0$, to emphasize the variation in width.  The four $\Delta x$ regions
yield 58\%, 25\%, 14\%, and 3\% of the total charge, with the largest
contributions coming from the bins with the  smallest values of  $x$. For $x
\approx 0.1$, the half-maximum width is 0.5~fm, while for $x \approx 0.8$, it
is 0.12~fm. Thus the large $x$ quarks (mainly $u$ quarks in the proton) play
an increasingly prominent role in the charge distribution at small values of
$b$. The curves shown in Fig.~\ref{fig:proton_scaled} are obtained using the
GPD of Ref.~\cite{diehl05}; The results obtained from the Guidal \textit{et
al}, parameterization for the GPDs are barely distinguishable. The GPDs
of Ref.\cite{ahmad07} also have a strong tendency to be constrained to smaller
and smaller values of $b$ as the value of $x$ increases.  We evaluate the GPDs
of all three models using the starting scale $Q_0^2$ of each model.

\begin{center}
 \begin{figure}[htb]
   \includegraphics[width=9.0cm]{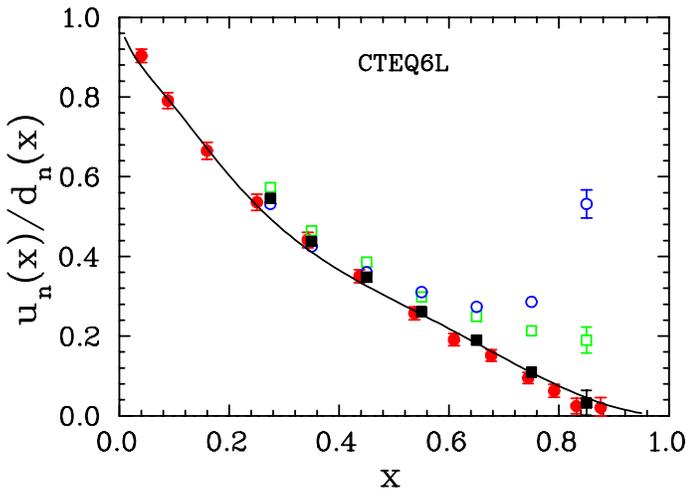}
   \caption{  
Ratio of u quarks to d quarks in the neutron from several analyses of
deuteron and proton data.  The solid line is the CTEQ6L
parameterization}
     \label{fig:ud_neutron}
 \end{figure}
\end{center}

\begin{center}
 \begin{figure}[htb]
   \includegraphics[width=8.0cm]{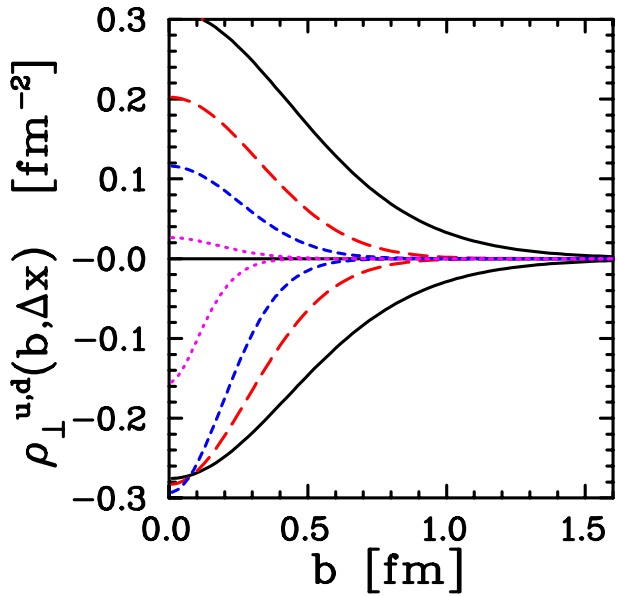}
   \caption{The $u$ and $d$ quark contributions to the neutron
transverse charge density, $\rho_\perp^u(b,\Delta x)$ and
$\rho_\perp^d(b,\Delta x)$. Here the quark flavor refers to the neutron 
($u$ in the proton is $d$ in the neutron).
The curves correspond to the same $\Delta x$
regions as in Fig.~\ref{fig:proton_scaled}. The largest contributions come
from small $x$, where $u$ and $d$ quarks contribute similar
amounts of charge. As one goes to larger $x$ values, the charge is shifted to
smaller values of $b$, while at the same time the up quark distribution
drops rapidly with respect to the down quarks. Reprinted with
   permission from Ref.~$^6$.        Copyright 2008 by the American Physical Society.}
   \label{fig:neutron_bins}
 \end{figure}
\end{center}

Taking what we have learned from the proton, we now consider the neutron. 
Figure~\ref{fig:ud_neutron} shows the ratio of the up- to down-quark
distributions in the neutron, as extracted from various analyses of deuteron
and proton data~\cite{whitlow92,melnitchouk96,arrington09} (using different
models for the nuclear corrections in dueterium), and from the
CTEQ6L~\cite{pumplin02} parameterization. For $x \to 0$, the up and down quark
distributions are similar, and the contribution to the charge distribution
from this limit should be similar to that of the proton; a broad distribution
of net positive charge.  For $x=0.3$ and above, the $u$ quark distribution is
less than half the $d$ quark distribution, yielding a net negative
contribution to the charge. Because the distribution of quarks is more
localized near $b=0$ as $x$ increases, a negative peak can be formed if there
is a sufficiently large contribution from down quarks at large $x$ values. 
Above $x=0.5$, $d_n(x)/u_n(x)$ is at least three, except for the density-based
extrapolation (blue circles), and continues to increase with $x$.  As
discussed in Ref.~\cite{arrington09}, even if one uses a density-based
extrapolation of the EMC effect to apply nuclear corrections, the
implementation used in this analysis significantly overestimates the effect
for deuterium, and yields an unrealistically large result at very large $x$
values. In this region, the net impact on the charge distribution will be
negative, and will be peaked at smaller values of $b$. This is shown in
Fig.~\ref{fig:neutron_bins}, which  separately shows the contributions to the
neutron charge density from $u$ and $d$ quarks based on the GPD of
Ref.~\cite{diehl05}.  The distributions of Refs.~\cite{guidal05,ahmad07} yield
somewhat different results, but  exhibit the same qualitative behavior. For
example, the GPDs of Ref.~\cite{diehl05}, shown in
Fig.~\ref{fig:neutron_bins}, yield a negative central neutron  charge density 
for values of $x$ between 0.15 and 0.3 and between 0.3 and 0.465, but for the
GPDs of Ref.~\cite{guidal05}, the central density is positive unless $x$ is
slightly greater than 0.465.

\begin{center}
 \begin{figure}[htb]
   \includegraphics[width=8.0cm]{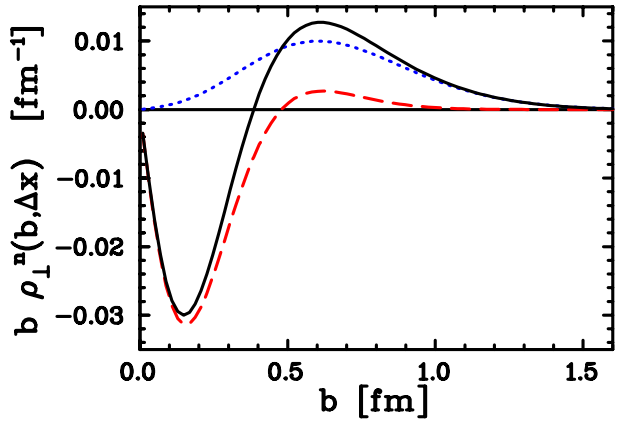}
   \caption{Transverse charge density for the neutron.  The
dotted line is the contribution from $x<0.23$, dashed is that for $x>0.23$, and
the solid is the total. Reprinted with permission from Ref.~$^6$. Copyright 2008 by the American Physical Society.}
   \label{fig:neutron_total}
 \end{figure}
\end{center}

Next we examine the total charge distribution of the neutron.
Fig.~\ref{fig:neutron_total} shows the charge distribution of the neutron,
separating out the contributions from low and high $x$
regions, and shows $b \rho_\perp^n$ to suppress the very large density at the
center. For $x<0.23$ the charge distribution (dotted line) is positive for
$b<1.5$~fm and slightly negative distribution at larger radii.  For $x>0.23$,
the contribution (dashed line) is largely negative, and highly localized below
0.5~fm. The negative region at the center of the neutron transverse charge
distribution arises a natural consequence of the model-independent definition
of the charge density.  The low momentum partons have a larger spatial extent
and reproduce the intuitive result of the pion cloud picture: a positive core
with a small negative tail at large distances, although the negative tail is
difficult to see for this parameterization of the GPD, given the large scale
required to show the negative core at small $b$.

\begin{center}
 \begin{figure}[htb]
   \includegraphics[width=8.0cm,height=4.5cm]{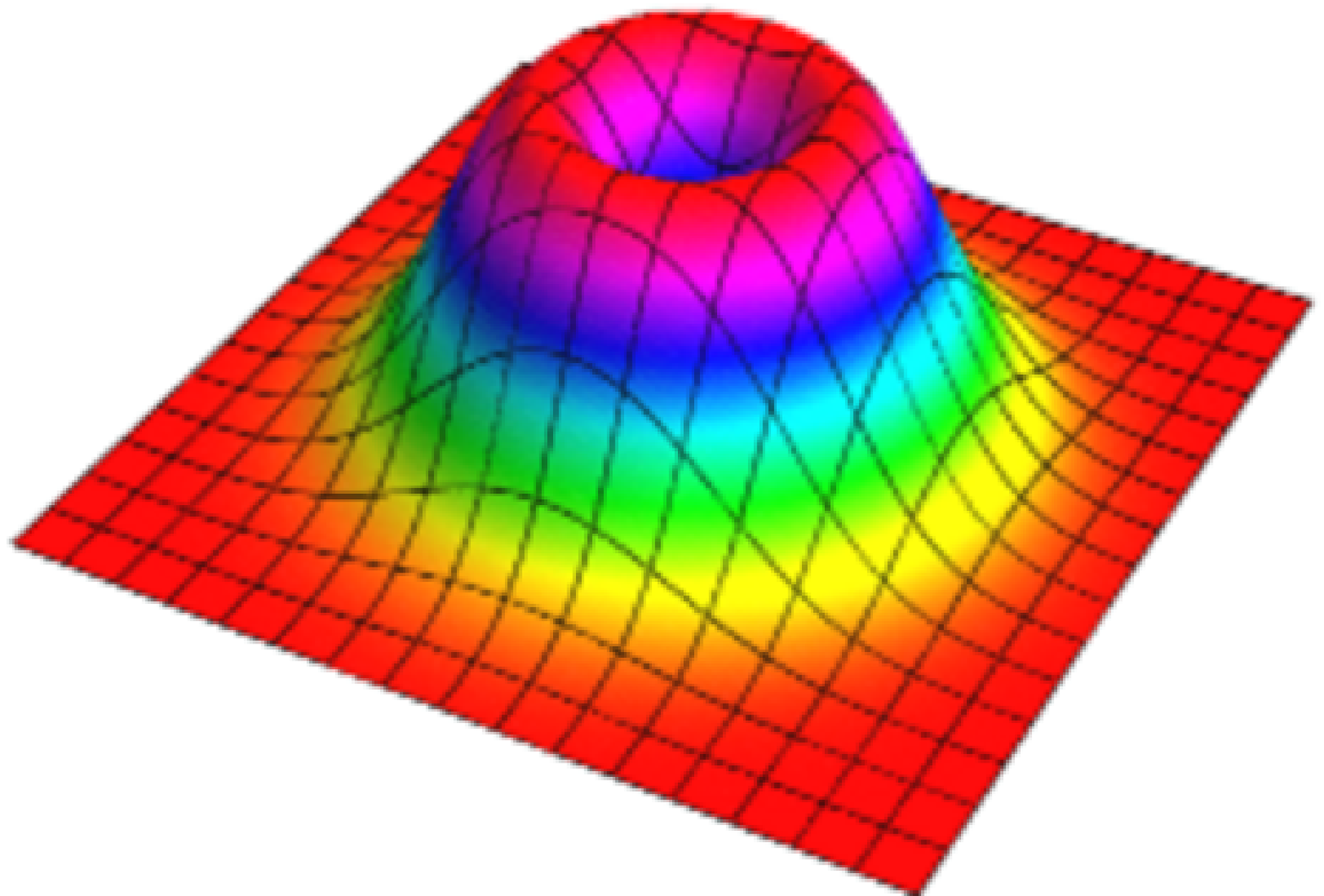}
   \includegraphics[width=8.0cm]{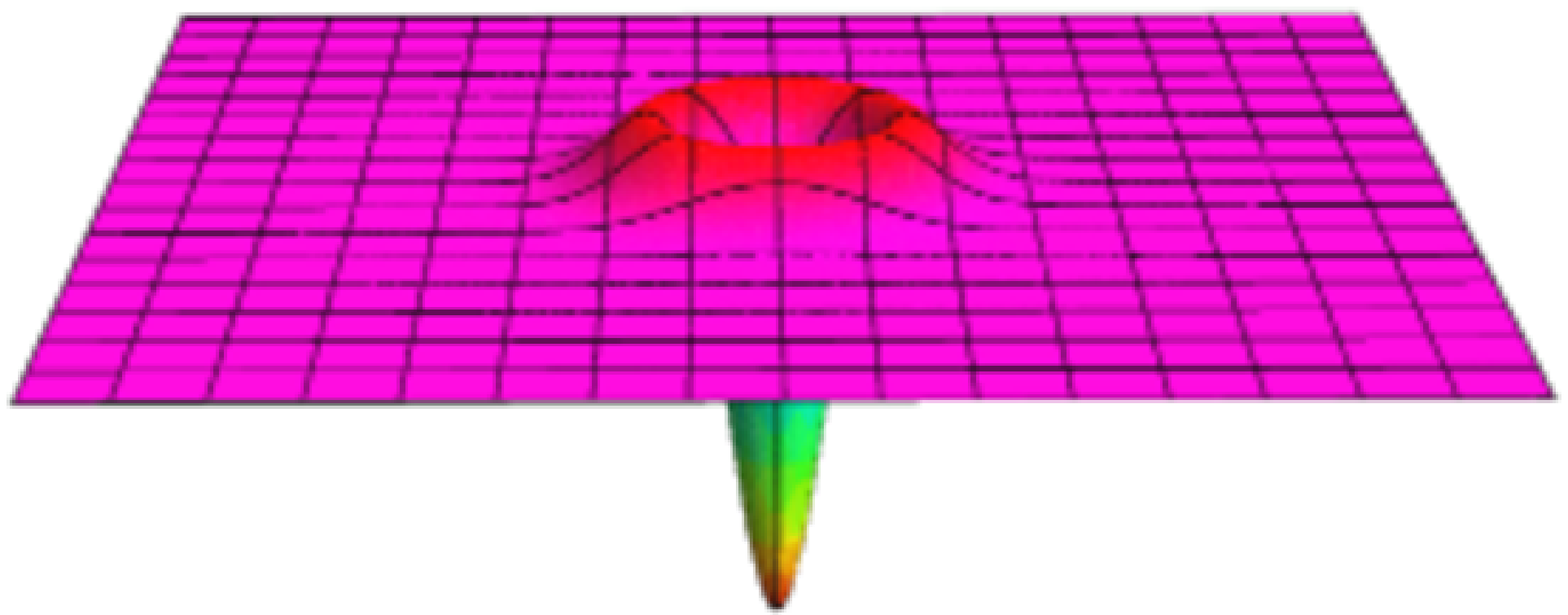}
   \includegraphics[width=8.0cm]{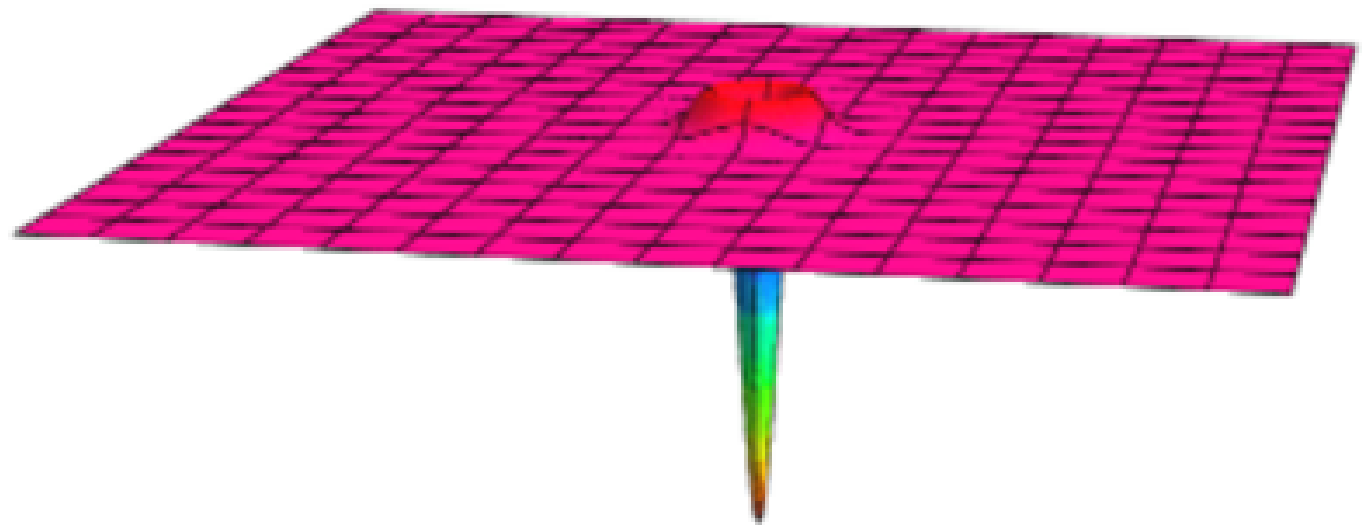}
   \caption{Neutron generalized parton distributions $\rho(b,x)$ 
 for $x=0.1$ (top), $x=0.3$
(middle), and $x=0.5$ (bottom).}
   \label{pt5}
 \end{figure}
\end{center}

We also study the quantity 
\bea \rho_\perp(b,x)\equiv \sum_q\;e_q\rho^q_\perp(b,x)\eea
to obtain a pictorial view of the transverse charge density for
specific values of $x$ (Fig.~\ref{pt5}). The
striking feature of the negative spike appears prominently for $x=0.3$
and more prominently for $x=0.5$. These figures  show how the central negative 
charge density appears more and more prominent as $x$
increases. Clearly the negatively charged $d$ quarks dominate at the
center of the neutron. 

\begin{center}
 \begin{figure}[htb]
   \includegraphics[width=8.0cm]{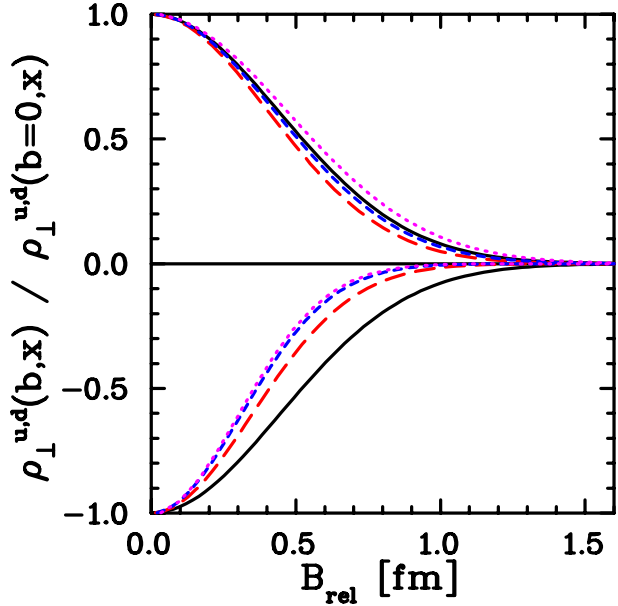}
   \caption{The $u$ and $d$ quark contributions to 
$\rho^{{\rm Spec},n}_\perp(B_{\rm rel},x)$ see \eq{redefine}. vs $B_{\rm rel}$
for $x=0.1$~(solid), 0.3~(long-dash), 0.5~(short-dash), and 0.7~(dotted). The
curves are scaled to unity at $B_{\rm rel}=0$. 
Here the quark flavor refers to the neutron 
($u$ in the proton is $d$ in the neutron). Reprinted with permission
   from Ref.~$^6$.
 Copyright 2008 by the American Physical Society.}
   \label{fig:neutron_spectator}
 \end{figure}
\end{center}

As discussed earlier, the longitudinal-momentum weighting used in determining
the nucleon center of mass leads to a strong correlation between the position
of the struck quark and the center of the nucleon for large $x$.  It is
informative to try and remove this effect to obtain something that is closer
to our intuitive picture.  We can do this by examining the position of the
struck quark relative to the center of the \textit{spectator} system, so that
the struck quark does not influence the definition of the center of the neutron.
This can be approximated by looking at the position of the struck quark
relative to the spectators.  We use \eq{tcm} 
with the origin set to the center of momentum, for a
struck quark at $(x_1,{\bf b_1})\equiv(x,\bfb)$, to  determine the
momentum-weighted spectator position, ${\bf b_{\rm spec}}$, and the relative
distance from the struck quark to the spectator quarks:
\bea
x_1 \bfb_1+\sum_{i>1}x_i\bfb_i=x\bfb+(1-x)\bfb_{\rm spec}=0,\\
{\bf B_{\rm rel}} = \bfb - {\bf b_{\rm spec}}=\frac{\bfb}{(1-x)}=\bf{B_{\rm rel}}.
\eea
We exhibit the dependence on ${\bf B_{\rm rel}}$ by defining a function
\bea
\rho^{\rm Spec}_\perp({\bf B}_{rel},x)\equiv \rho_\perp(\bfB_{rel}(1-x),x)\label{redefine}
\eea
which gives the probability that a struck quark of longitudinal momentum
fraction $x$ is a distance ${\bf B_{\rm rel}}$ away from the spectator center of
momentum.  Figure~\ref{fig:neutron_spectator} shows this rescaled version of
$\rho_\perp(b)$, with the contribution at each $x$ value normalized to
unity at $b=0$.  
The quantity $\rho^{\rm Spec}_\perp({\bf B}_{\rm rel},x)$ 
does not correspond to a true density, but
can provide a better approximation to our intuitive picture of the charge
distribution, as it removes the influence of the struck quark on defining the
center of the nucleon. While the charge distribution coming from very low $x$
quarks has a greater spatial extent, the decreasing width of the
$\rho_\perp(b)$ distribution for large $x$ quarks is essentially completely
removed when looking at $B_{\rm rel}$.

We summarize our findings with the statement that, using the model GPDs of
Refs.~\cite{diehl05,guidal05,ahmad07}, the dominance of the neutron's $d$
quarks at high values of $x$ leads to a negative contribution to the charge
density which, due to the definition of $\bfb$, becomes localized near the
center of mass of the neutron.  This localization does not appear when
examined as a function of the position of the struck quark relative to the
spectators, and is an consequence of the fact that quarks with a large
longitudinal momentum play an important role in defining the transverse
position of the neutron.

\section*{Acknowledgments}

This work was supported by the U. S. Department of Energy, Office of Nuclear
Physics, under contracts FG02-97ER41014 and DE-AC02-06CH11357.  We thank
D.~Geesaman, R.~Holt, P.~Kroll, C.~Roberts, M. Vanderhaeghen,
 and B.~Wojtsekhowski for useful
discussions.  We thank the ECT* for hosting a workshop where many of the
calculations we present were performed.



\begin{thebibliography}{0}


\bibitem{gao03}  H.~Gao,
  Int.\ J.\ Mod.\ Phys.\  E {\bf 12} (2003) 1
  [Erratum-ibid.\  E {\bf 12} (2003) 567].

\bibitem{hydewright04} C.~E.~Hyde-Wright and K.~de Jager,
  Ann.\ Rev.\ Nucl.\ Part.\ Sci.\  {\bf 54}, (2004)  21.

\bibitem{ perdrisat07}C.~F.~Perdrisat, V.~Punjabi and M.~Vanderhaeghen,
  Prog.\ Part.\ Nucl.\ Phys.\  {\bf 59} (2007) 694.

 \bibitem{arrington07a}  J.~Arrington, C.~D.~Roberts and J.~M.~Zanotti,
  J.\ Phys.\ G {\bf 34} (2007) S23

\bibitem{miller07}
G.~A.~Miller,
  Phys.\ Rev.\ Lett.\  {\bf 99} (2007) 112001.

\bibitem{ma08} G.~A.~Miller and J.~Arrington,
  Phys.\ Rev.\  C {\bf 78} (2008) 032201.
\bibitem{soper} D.E. Soper, Phys.\ Rev.\ D\ {\bf 5} 
 (1972) 1956;
J. Kogut and D.E. Soper, 
Phys.\ Rev.\ D\ {\bf 1}  (1970) 2901.
\bibitem{soper77} D.~E.~Soper,
  Phys.\ Rev.\  D {\bf 15} (1977) 1141.

\bibitem{mbimpact} M.~Burkardt,
  Int.\ J.\ Mod.\ Phys.\  A {\bf 18} (2003) 173.
\bibitem{diehl2} 
 M.~Diehl,
  Eur.\ Phys.\ J.\  C {\bf 25}  (2002)223,
  [Erratum-ibid.\  C {\bf 31} (2003) 277].


\bibitem{miller07b}
  G.~A.~Miller, E.~Piasetzky and G.~Ron,
  Phys.\ Rev.\ Lett.\  {\bf 101} (2008) 082002.
\bibitem{carlson08}
  C.~E.~Carlson and M.~Vanderhaeghen,
  Phys.\ Rev.\ Lett.\  {\bf 100}, 032004 (2008)


\bibitem{Bradford:2006yz}R.~Bradford, A.~Bodek, H.~Budd and J.~Arrington,
  Nucl.\ Phys.\ Proc.\ Suppl.\  {\bf 159} (2006) 127


\bibitem{Kelly:2004hm}J.~J.~Kelly,
  Phys.\ Rev.\  C {\bf 70} (2004) 068202.
\bibitem{Arrington:2003qk}
 J.~Arrington,
  Phys.\ Rev.\  C {\bf 69}, 022201 (R) (2004);


\bibitem{cbm}  A.~W.~Thomas, S.~Th\'eberge and G.~A.~Miller,
  Phys.\ Rev.\  D {\bf 24}  (1981) 216;
G.~A.~Miller,
  Phys.\ Rev.\  C {\bf 66} (2002) 032201.
\bibitem{isgurk}
J.~L.~Friar,
  Part.\ Nucl.\  {\bf 4}, 153 (1972);
R.~D.~Carlitz, S.~D.~Ellis and R.~Savit,
  Phys.\ Lett.\  B {\bf 68}, 443 (1977);
  N.~Isgur, G.~Karl and D.~W.~L.~Sprung,
  Phys.\ Rev.\  D {\bf 23}, 163 (1981).

\bibitem{ji96} X.~D.~Ji,
  Phys.\ Rev.\  D {\bf 55} (1997) 7114

\bibitem{radyushkin97}  A.~V.~Radyushkin,
  Phys.\ Rev.\  D {\bf 56} (1997) 5524


\bibitem{burkardt00} M.~Burkardt,
  Phys.\ Rev.\  D {\bf 62} (2000) 071503
  [Erratum-ibid.\  D {\bf 66} (2002) 119903]

\bibitem{guidal05}  M.~Guidal, M.~V.~Polyakov, A.~V.~Radyushkin and M.~Vanderhaeghen,
  Phys.\ Rev.\  D {\bf 72} (2005) 054013
\bibitem{diehl05} M.~Diehl, T.~Feldmann, R.~Jakob and P.~Kroll,
  Eur.\ Phys.\ J.\  C {\bf 39}, 1 (2005)
\bibitem{ahmad07}  S.~Ahmad, H.~Honkanen, S.~Liuti and S.~K.~Taneja,  
  Phys.\ Rev.\  D {\bf 75}, 094003 (2007)
\bibitem{tiburzi04}   B.~C.~Tiburzi, W.~Detmold and G.~A.~Miller,
  Phys.\ Rev.\  D {\bf 70}, 093008 (2004)


\bibitem{DYW} S.~D.~Drell and T.~M.~Yan,
  Phys.\ Rev.\ Lett.\  {\bf 24}, 181 (1970);
 G.~B.~West,
  Phys.\ Rev.\ Lett.\  {\bf 24}, 1206 (1970).


\bibitem{pumplin02}J.~Pumplin, D.~R.~Stump, J.~Huston, H.~L.~Lai, P.~M.~Nadolsky and W.~K.~Tung,
  JHEP {\bf 0207}, 012 (2002)
\bibitem{miller90} G.~A.~Miller, B.~M.~K.~Nefkens and I.~Slaus,
  Phys.\ Rept.\  {\bf 194}, 1 (1990).
\bibitem{miller98}G.~A.~Miller,
  Phys.\ Rev.\  C {\bf 57}, 1492 (1998)
\bibitem{londergan98} J.~T.~Londergan and A.~W.~Thomas,
  Prog.\ Part.\ Nucl.\ Phys.\  {\bf 41}, 49 (1998)
\bibitem{miller06} G.~A.~Miller, A.~K.~Opper and E.~J.~Stephenson,
  Ann.\ Rev.\ Nucl.\ Part.\ Sci.\  {\bf 56}, 253 (2006)

\bibitem{whitlow92}
  L.~W.~Whitlow, E.~M.~Riordan, S.~Dasu, S.~Rock and A.~Bodek,
  Phys.\ Lett.\  B {\bf 282}, 475 (1992).

\bibitem{melnitchouk96}
  W.~Melnitchouk and A.~W.~Thomas,
  Phys.\ Lett.\  B {\bf 377}, 11 (1996)

\bibitem{arrington09}
  J.~Arrington, F.~Coester, R.~J.~Holt and T.~S.~Lee,
  J.\ Phys.\ G {\bf 36}, 025005 (2009)

\end{thebibliography}
\end{document}